\documentclass[aps,pre,preprint,groupedaddress,showpacs]{revtex4}
\usepackage{graphicx}
\begin{document}
\newcommand{\ud}{{\mathrm d}}
\newcommand{\umod}{\mathrm{mod}}
\newcommand{\sech}{\mathrm{sech}}
% Use the \preprint command to place your local institutional report
% number in the upper righthand corner of the title page in preprint mode.
% Multiple \preprint commands are allowed.
% Use the 'preprintnumbers' class option to override journal defaults
% to display numbers if necessary
%\preprint{}

%Title of paper
\title{Subthreshold Stochastic Resonance: Rectangular signals can cause anomalous large gains }

% repeat the \author .. \affiliation  etc. as needed
% \email, \thanks, \homepage, \altaffiliation all apply to the current
% author. Explanatory text should go in the []'s, actual e-mail
% address or url should go in the {}'s for \email and \homepage.
% Please use the appropriate macro foreach each type of information

% \affiliation command applies to all authors since the last
% \affiliation command. The \affiliation command should follow the
% other information
% \affiliation can be followed by \email, \homepage, \thanks as well.
\author{Jes\'us Casado-Pascual}
\email[]{jcasado@us.es}
\homepage[]{http://numerix.us.es}
\author{Jos\'e  G\'omez-Ord\'o\~nez}
\author{ Manuel Morillo}
\affiliation{F\'{\i}sica Te\'orica,
Universidad de Sevilla, Apartado de Correos 1065, Sevilla 41080, Spain}

%\thanks{}
%\altaffiliation{}

\author{Peter H\"anggi}
\affiliation{Institut f\"ur Physik,
Universit\"at Augsburg, Universit\"atsstra\ss e 1, D-86135 Augsburg,
Germany}

%Collaboration name if desired (requires use of superscriptaddress
%option in \documentclass). \noaffiliation is required (may also be
%used with the \author command).
%\collaboration can be followed by \email, \homepage, \thanks as well.
%\collaboration{}
%\noaffiliation

\date{\today}

\begin{abstract}
The main objective of this work is to explore aspects of stochastic
resonance (SR) in noisy bistable, symmetric systems driven by
subthreshold periodic rectangular external signals possessing a {\em
large} duty cycle of unity. Using a precise numerical solution of the
Langevin equation, we carry out a detailed analysis of the behavior of
the first two cumulant averages, the correlation function and its
coherent and incoherent parts. We also depict the non-monotonic behavior
versus the noise strength of several SR quantifiers such as the average
output amplitude, i.e. the spectral amplification (SPA), the
signal-to-noise ratio (SNR) and the SR-gain.  In particular, we find
that with {\em subthreshold} amplitudes and for an appropriate duration
of the pulses of the driving force the phenomenon of stochastic
resonance (SR), is accompanied by SR-gains exceeding unity.  This analysis
thus sheds new light onto the interplay between nonlinearity and the
nonlinear response which in turn yields nontrivial, unexpected SR-gains
above unity.
\end{abstract}

% insert suggested PACS numbers in braces on next line
% insert suggested keywords - APS authors don't need to do this
%\keywords{}

%\maketitle must follow title, authors, abstract, \pacs, and \keywords
\maketitle
\section{Introduction}
Most of the studies on the phenomenon of Stochastic Resonance (SR) in
dynamical systems have been devoted to systems driven by sinusoidal
terms (see \cite{PT96,RMP,Wiesenfeld98,Anish99,Chemphyschem02} for
reviews). Several analytical approximations have been put forward to
explain SR. In the approach of McNamara and Wiesenfeld \cite{McNWie89},
the Langevin dynamics is replaced by a reduced two-state model that
neglects the intra-well dynamics. The general ideas of Linear Response
Theory (LRT) have been applied to situations where the input amplitude
is small \cite{Hanggi78,PR82,RMP,EPL89,GM89,PRA91}.  In
\cite{EPL89,PRA91} the Floquet theory has been applied to the
corresponding Fokker-Planck description.  For very low input
frequencies, an adiabatic ansatz has been invoked \cite{PRA91}. Even
though these alternative analytical approaches provide an explanation of
SR for different regions of parameter values, their precise limits of
validity remain to be determined. In recent work,
\cite{EPLcasado,FNLcasado} we have explored the validity of LRT for
sinusoidal and multifrequency input signals with low frequency. Our
results indicate a breakdown of the LRT description of the average
behavior for low frequency, subthreshold amplitude inputs.

Several quantifiers have been used to characterize SR in noisy,
continuous, systems. The average output amplitude, or the spectral
amplification (SPA), has been studied in Refs.~\cite{EPL89,PRA91} and
the phase of the output average in Refs.  \cite{dykman92,junhan93},
respectively. Those parameters as well as the signal-to-noise ratio
(SNR) \cite{McNWie89}, exhibit a non-monotonic behavior with the noise
strength which is representative of SR.  An important quantity is the
SR-gain, defined as the ratio of the SNR of the output over the input
SNR. It has been repeatedly pointed out that the SR-gain can not exceed
unity as long as the system operates in a regime described by LRT
\cite{dewbia95,casgom03}. Beyond LRT there exists no physical reason
that prevents the SR-gain to be larger than $1$, as it has been
demonstrated in \cite{haninc00} for super-threshold sinusoidal input
signals, and in analog experiments in \cite{ginmak00,ginmak01} for
subthreshold input signals with many Fourier components and a small
duty cycle \cite{casgom03}.

\begin{figure}
\includegraphics[width=10cm]{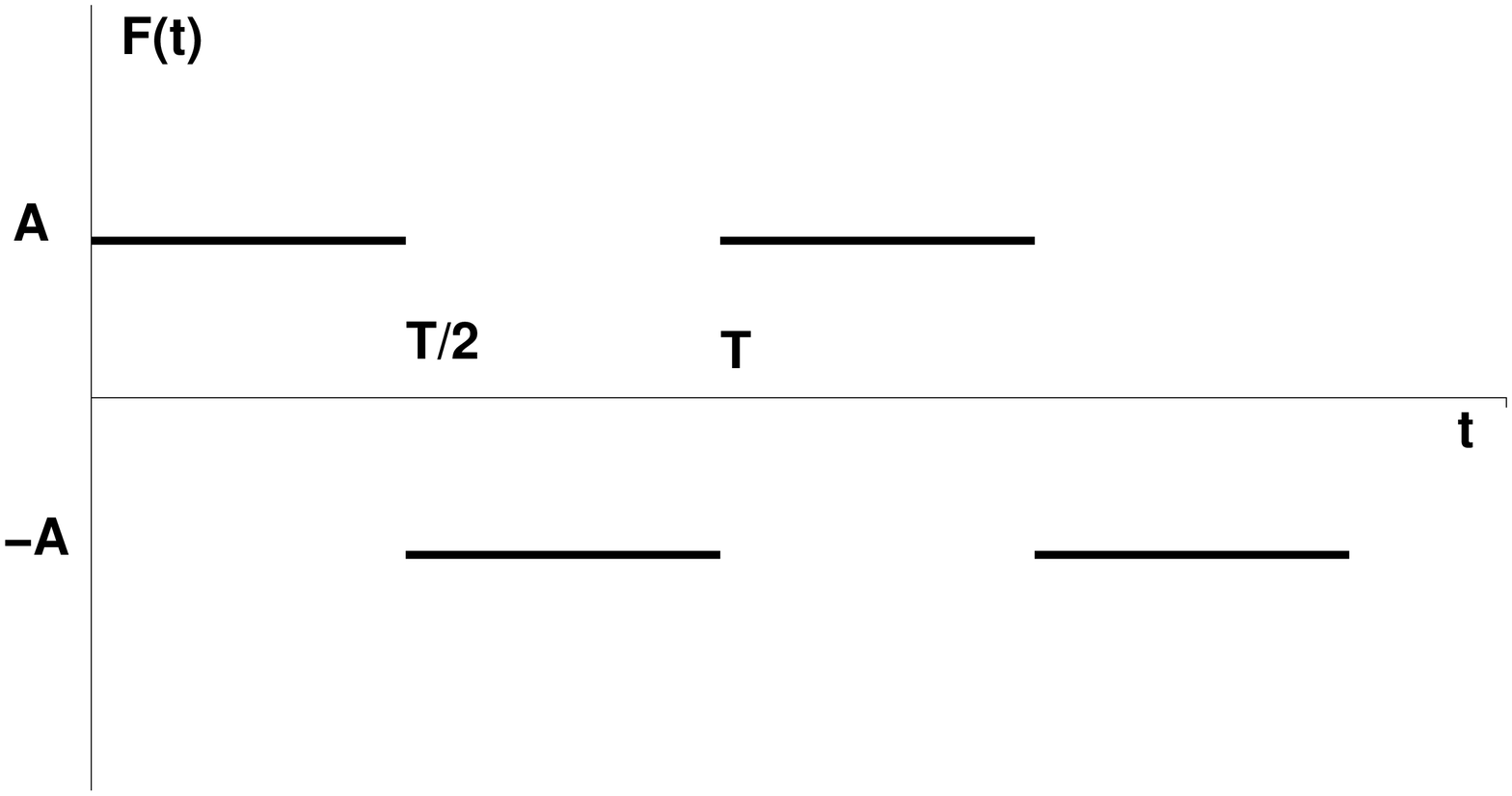}
\caption{\label{fuerza}Sketch of a rectangular periodic signal with
duty cycle 1, amplitude $A$ and period $T$}
\end{figure}

In this work, we will make use of numerical solutions of the Langevin
equation following the methodology in \cite{casgom03}, to analyze SR in
noisy bistable systems, driven by periodic piecewise constant signal
with two amplitude values of opposite signs (rectangular signal) (see
Fig.~ \ref{fuerza}).  There
are several relevant time scales in the dynamics of these systems: i)
$t_{asym}$, the time interval within each half period of the driving force,
during which the diffusing particle sees an asymmetric constant two-well
potential; ii) $t_{inter}$, the time scale associated with the inter-well
transitions in both directions; and iii) $t_{intra}$, the time scale
associated to intra-well dynamics. The inter-well and intra-well time
scales depend basically on the noise strength $D$ and the amplitude of
the driving term. The dependence of these two time scales with those
parameters is certainly very different, being more pronounced for
$t_{inter}$. Typically, for the range of parameter values associated
with SR, the intra-well time scale is shorter than the inter-well one.

We will evaluate the long-time average behavior of the output and the
second cumulant. These two quantities were studied some time ago by two
of us \cite{Morillo95} for periodic rectangular driving signals. Here,
we will further extend our work to the analysis of the correlation
function and its coherent and incoherent parts. The knowledge of all
these quantities provides a very useful information for the explanation
of SR as indicated by the non-monotonic behavior with the noise strength
of the output amplitude and the SNR. In particular, the knowledge
of the incoherent part of the correlation function is of outmost
importance for a correct determination of the SNR. Furthermore, for a
given subthreshold amplitude, we will demonstrate that, if there exist a
range of noise values such that $t_{inter}$ is shorter than
$t_{asym}$, then it is possible to observe stochastic amplification and,
simultaneously, SR-gains larger than unity. This is strictly
forbidden by linear response theory, as we have recently
shown \cite{casgom03}. Thus, a simultaneous appearance of stochastic
amplification and SR-gains above $1$ implies a strong violation of linear
response theory.

\section{Model system and SR quantifiers}
Let us consider a system characterized by a single degree of freedom,
$x$, subject to the action of a zero average Gaussian white noise with
$\langle \xi(t)\xi(s)\rangle = 2D\delta(t-s)$ and driven by an
external periodic signal $F(t)$ with period $T$.  In the Langevin
description, its dynamics is generated by the equation
\begin{equation}
\label{langev} \dot{x}(t)=-U'\left[ x(t) \right]+F(t)+\xi(t).
\end{equation}
The corresponding linear Fokker-Planck equation (FPE) for the
probability density $P(x,t)$ reads
\begin{equation}
\label{FP}
\frac{\partial}{\partial t}P(x,t)={\hat {\mathcal L}}(t) P(x,t),
\end{equation}
where
\begin{equation}
{\hat {\mathcal L}}(t)=\frac{\partial}{\partial x}\left[ U'(x)-F(t)+ D
\frac{\partial}{\partial x}\right].
\end{equation}
In the expressions above, $U'(x)$ represents the derivative of the
potential $U(x)$. In this work, we will consider a bistable potential
$U(x)=-x^2/2+x^4/4$. The periodicity of the external driving $F(t)$
allows its Fourier series expansion in the harmonics of the
fundamental frequency $\Omega=2\pi/T$, i.e.,
\begin{equation}
\label{fourf}
F(t)=\sum_{n=1}^\infty \left [ f_n \cos (n \Omega t) + g_n \sin
(n \Omega t) \right ],
\end{equation}
with the Fourier coefficients, $f_n$ and $g_n$, given by
\begin{eqnarray}
f_n&=&\frac 2T \, \int_0^T \ud t\, F(t) \cos (n \Omega t),
\nonumber \\
g_n&=&\frac 2T \, \int_0^T \ud t\, F(t) \sin (n \Omega t).
\end{eqnarray}
Here, we are assuming that the cycle average of the external driving
over its period equals zero. In this work, we will focus our attention
to multi-frequency input forces with ``rectangular'' shape given by
\begin{equation}
\label{pulse}
F(t)= \left \{ \begin{array}
{r@{\quad:\quad}l}
A & 0\le t < \frac T2 \\
-A& \frac T2 \le t < T
\end{array}
\right .
\end{equation}
as sketched in Fig.~\ref{fuerza}. The external force remains constant
at a value $A$ during each half-period and changes sign for the second
half of the period.  The duty cycle of the signal is defined by the time
span the signal is nonzero over the total period of the signal; thus,
the rectangular signal in Eq.\ (\ref{pulse}) consequently possesses a
duty cycle of unity.

The two-time correlation function $\langle x(t+\tau)
x(t)\rangle_{\infty}$ in the limit $t \rightarrow \infty$ is given
by
\begin{equation}
\langle x(t+\tau) x(t)\rangle_{\infty} =\int_{-\infty}^{\infty} \ud x'
\,x' P_{\infty}(x',t) \int_{-\infty}^{\infty}
\ud x \,x P_{1|1}(x,t+\tau|x',t),
\end{equation}
where $P_{\infty}(x,t)$ is the time-periodic, asymptotic long time
solution of the FPE and the quantity $P_{1|1}(x,t+\tau|x',t)$
denotes the two-time conditional probability density that the
stochastic variable will have a value near $x$ at time $t+\tau$ if
its value at time $t$ was exactly $x'$. It can been shown
\cite{RMP,PRA91} that, in the limit $t \rightarrow \infty$, the
two-time correlation function $\langle x(t+\tau)
x(t)\rangle_\infty$ becomes a periodic function of $t$ with the
period of the external driving. Then, we define the one-time
correlation function, $C(\tau)$, as the average of the two-time
correlation function over a period of the external driving, i.e.,
\begin{equation}
\label{ctau}
C(\tau)= \frac{1}{T} \int_{0}^{T} \ud t \, \langle x(t+\tau)
x(t)\rangle_{\infty}.
\end{equation}
The correlation function $C(\tau)$ can be written exactly as the
sum of two contributions: a coherent part, $C_{coh}(\tau)$, which
is periodic in $\tau$ with period $T$, and an incoherent part
which decays to $0$ for large values of $\tau$. The coherent part
$C_{coh}(\tau)$ is given by \cite{RMP,PRA91}
\begin{equation}
\label{chtau}
C_{coh}(\tau) =\frac{1}{T} \int_{0}^{T} \ud t \,\langle x(t+\tau)
\rangle_{\infty} \langle x(t) \rangle_{\infty},
\end{equation}
where $\langle x(t) \rangle_{\infty}$ is the average value evaluated
with the asymptotic form of the probability density, $P_{\infty}(x,t)$.

According to McNamara and Wiesenfeld \cite{McNWie89}, the output SNR is
defined in terms of the Fourier transform of the coherent and
incoherent parts of $C(\tau)$. As the correlation function is even
in time and we evaluate its time dependence for $\tau \ge 0$, it
is convenient to use its Fourier cosine transform, defined as
\begin{equation}
\label{fourier}
\tilde{C}(\omega)=\frac 2\pi \int_0^\infty \ud\tau\,C(\tau) \cos (\omega
\tau);\; C(\tau)=\int_0^\infty \ud\omega\, \tilde{C}(\omega) \cos (\omega
\tau).
\end{equation}
The value of the output SNR is then obtained from:
\begin{equation}
\label{snr}
R_{out} =\frac {\lim_{\epsilon \rightarrow 0^+}
\int_{\Omega-\epsilon}^{\Omega+\epsilon} \ud\omega\;
\tilde{C}(\omega)}{\tilde{C}_{incoh}(\Omega)}.
\end{equation}
Note that this definition of the SNR differs by a factor $2$, stemming
from the same contribution at $\omega = - \Omega$, from the
definitions used in earlier works \cite {RMP,PRA91}. The periodicity of
the coherent part gives rise to delta peaks in the spectrum. Thus, the
only contribution to the numerator in Eq.\ (\ref{snr}) stems from the
coherent part of the correlation function. The evaluation of the SNR
requires the knowledge of the Fourier components of $C_{coh}(\tau)$ and
$C_{incoh}(\tau)$ at the fundamental frequency of the driving
force. Thus, rather than knowledge of the entire Fourier spectrum, only
two well defined numerical quadratures are needed. These are:
\begin{equation}
\label{snr1}
R_{out}=\frac{Q_u}{Q_l},
\end{equation}
where
\begin{equation}
\label{num}
Q_u= {\frac 2T} \int_0^T \ud \tau \,C_{coh}(\tau) \cos (\Omega
\tau),
\end{equation}
and
\begin{equation}
\label{den}
Q_l=\frac 2\pi \int_0^\infty \ud \tau \,C_{incoh}(\tau) \cos (\Omega
\tau ).
\end{equation}
The SNR for an input signal $F(t)+\xi(t)$ is given by
\begin{equation}
\label{snrinp}
R_{inp}=\frac{ \pi(f_1^2+g_1^2)}{4D}.
\end{equation}
The SR-gain $G$ is consequently defined as the ratio of the SNR of the
output over the SNR of the input, namely,
\begin{equation}
\label{gain}
G=\frac {R_{out}}{R_{inp}}.
\end{equation}
\section{Numerical solution}
Stochastic trajectories, $x^{(j)}(t)$, are generated by numerically
integrating the Langevin equation [Eq.\ (\ref{langev})] for every
realization $j$ of the white noise $\xi(t)$, starting from a given
initial condition $x_0$.  The numerical solution is based on the
algorithm developed by Greenside and Helfand \cite{hel79,grehel81} (consult
also the Appendix in Ref.~\cite{casgom03}).  After allowing for a relaxation
transient stage, we start recording the time evolution of each random
trajectory for many different trajectories. Then, we construct the long
time average value,
\begin{equation}
\label{avgnum}
\langle x(t) \rangle_\infty = \frac 1N \sum_{j=1}^N x^{(j)}(t),
\end{equation}
and the second cumulant,
\begin{equation}
\label{cumul}
\langle x^2(t)\rangle_\infty-\langle x(t)\rangle^2_\infty =\frac 1N
\sum_{j=1}^N \left [x^{(j)}(t)\right ]^2 - \left [ \frac 1N \sum_{j=1}^N
x^{(j)}(t) \right ] ^2,
\end{equation}
where $N$ is the number of stochastic trajectories considered.
We also evaluate the two-time ($t$ and $\tau$) correlation function,
i.e.,
\begin{equation}
\langle x(t+\tau) x(t) \rangle_\infty = \frac 1N \sum_{j=1}^N
x^{(j)}(t+\tau) x^{(j)}(t),
\end{equation}
as well as the product of the averages
\begin{equation}
\langle x(t+\tau)\rangle_\infty \langle x(t) \rangle_\infty = \left [\frac 1N
\sum_{j=1}^N  x^{(j)}(t+\tau)  \right ] \left [\frac 1N
\sum_{j=1}^N x^{(j)}(t)  \right ].
\end{equation}
 The correlation function $C(\tau)$ and its coherent part
$C_{coh}(\tau)$ are then obtained using their definitions in
Eqs.~(\ref{ctau}) and (\ref{chtau}), performing the cycle average over
one period of $t$.  The difference between the values of $C(\tau)$ and
$C_{coh}(\tau)$ allows us to obtain the values for $C_{incoh}(\tau)$. It
is then straightforward to evaluate the Fourier component of
$C_{coh}(\tau)$ and the Fourier transform of $C_{incoh}(\tau)$ at the
driving frequency by numerical quadrature. With that information, the
numerator and the denominator for the output SNR [cf. Eqs.\
(\ref{snr1}), (\ref{num}) and (\ref{den})], as well as the SR-gain
[cf. Eq.\ (\ref{gain})], are obtained.
\section{Results}
\subsection{Response to a rectangular driving force with fundamental
frequency $\Omega=0.01$}
Consider an external driving of the type sketched in Fig.~\ref{fuerza}
with parameter values $\Omega=0.01$,
$A=0.25$. This amplitude is well below its threshold value defined,
for each driving frequency, as the minimum amplitude that can induce
repeated transitions between the minima of $U(x)$ in the absence of
noise. For the input considered here, the threshold amplitude is $A_T
\approx 0.37$. Note that this threshold value for the amplitude increases
with increasing driving frequency.

\begin{figure}
\includegraphics[width=10cm]{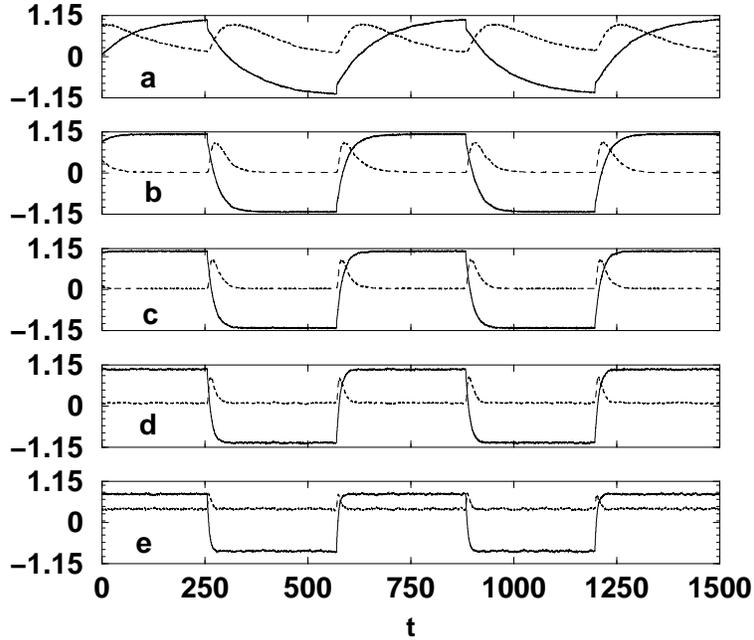}
\caption{\label{wp1} Time behavior of the average $\langle x(t)\rangle
_\infty$ (solid lines) and the second cumulant $\langle x^2(t)
\rangle_\infty - \langle x(t)\rangle _\infty^2$ (dashed lines) for a rectangular
driving force with duty cycle 1, fundamental frequency $\Omega=0.01$ and
subthreshold amplitude $A=0.25$ for several values of the noise
strength: $D=0.02$ (panel a), $D=0.04$ (panel b), $D=0.06$ (panel c),
$D=0.1$ (panel d), $D=0.2$ (panel e). Notice that
due to the transients, $t=0$ in the graphs,  does not necessarily
coincide with the start of an external cycle.}
\end{figure}

In Fig.~\ref{wp1} we depict with several panels the behavior of the
first two cumulants, $ \langle x(t) \rangle_\infty$ and $\langle
x^2(t)\rangle_\infty -\langle x(t)\rangle^2_\infty$, for several
representative values of $D$ [from top to bottom $D=0.02$ (panel a),
$D=0.04$ (panel b), $D=0.06$ (panel c), $D=0.1$ (panel d) and $D=0.2$
(panel e)].  Notice that due to the transients, the time at which we
start recording data, $t=0$ in the graphs, does not necessarily coincide
with the start of an external cycle. The average is periodic with the
period of the driving force, while the second cumulant, due to the
reflection symmetry of the potential \cite{EPLcasado}, is periodic with
a period half of the period of the forcing term.

Next we consider the case of small noise intensity $D$ (say, $D=0.02$ as
in panel a). The noise induces jumps between the wells. In each random
trajectory, a jump between the wells has a very short duration, but the
instants of time at which they take place for the different stochastic
trajectories are randomly distributed during a half-cycle. At this small
noise strength, the jumps are basically towards the lowest
minimum. Thus, because of this statistical effect, the average behavior
depicts the smooth evolution depicted in panel a of Fig.~\ref{wp1},
without sudden transitions between the wells. The evolution of the
second cumulant adds relevant information. The fact that it is rather
large during most of a period indicates that the probability density
$P_\infty(x,t)$ is basically bimodal during most of the external
cycle. It is only during very short time intervals around each
half-period that the probability distribution becomes monomodal around
one of the minima, and, consequently, the second cumulant is small. The
bimodal character arises from the fact that the noise is so small in
comparison with the barrier heights that jumps over the barrier are
rather infrequent during each half a period.

As the noise strength increases, the time evolution of $ \langle
x(t)\rangle_\infty$ follows closely the shape of the external force,
(cf. see panels b, c, d in Fig.~\ref{wp1} for $0.04 \le D \le 0.1$). This
behavior indicates that, for these parameter values, the jumps in the
different random trajectories are concentrated within short time
intervals around the instants of time at which the driving force
switches sign. The second cumulant remains very small during most of a
period, except for short time intervals around the switching times of
the external driver. Thus, for these intermediate values of $D$, the
probability distribution, $P_\infty(x,t)$ is basically monomodal, except
for small time intervals around the switching instants of time of the
periodic driver. Finally, as the noise strength is further increased ($D
> 0.1$), the probability distribution remains very broad most of the
time. Even though a large majority of random trajectories will still
jump over the barrier in synchrony with the switching times of the
driver, the noise is so large in comparison with the barrier heights,
that the probability of crossings over the barrier in both directions
can not be neglected at any time during each half cycle. The probability
distribution remains bimodal during a whole period, but asymmetric: the
larger fraction of the probability accumulates around the corresponding
lower minima of the potential during each cycle. Therefore, the average
output amplitude decreases, while the second cumulant depicts plateaus at
higher values than for smaller noise strengths (compare panels e and c
in Fig.~\ref{wp1}).

\begin{figure}
\includegraphics[width=10cm]{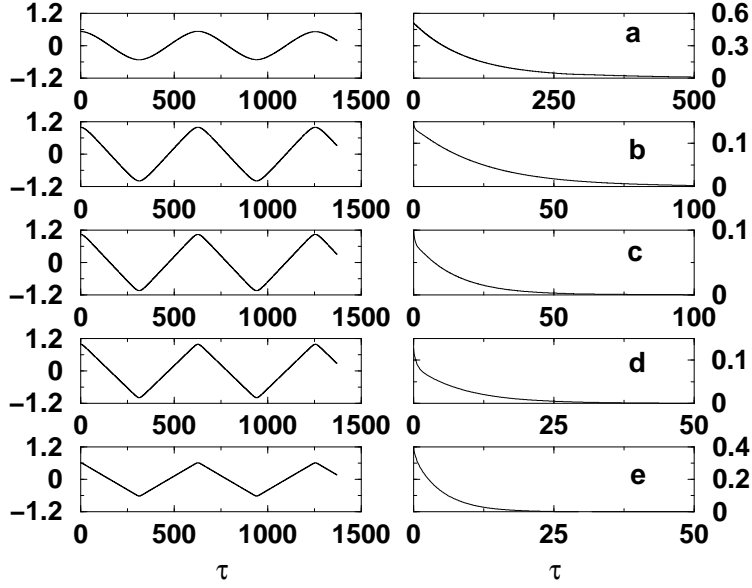}
\caption{\label{wp2} Time behavior of $C_{coh}(\tau)$ (left panels) and
$C_{incoh}(\tau)$ (right panels) for a rectangular driving force with duty
cycle 1, fundamental frequency $\Omega=0.01$ and subthreshold
amplitude $A=0.25$ for several values of the noise strength: $D=0.02$
(panel a), $D=0.04$ (panel b), $D=0.06$ (panel c), $D=0.1$ (panel d),
$D=0.2$ (panel e).}
\end{figure}

In Fig.~\ref{wp2}, we plot the coherent (left panels), $C_{coh}(\tau)$,
and incoherent (right panels), $C_{incoh}(\tau)$, components of the
correlation function $C(\tau)$ for the same parameter values as in
Fig.~\ref{wp1}. The coherent part shows oscillations with a period equal
to that of the driving force. Its shape changes with $D$. The amplitude
of the coherent part does not grow monotonically with $D$. Rather, it
maximizes at $D\approx 0.06$, which is consistent with the observed
behavior of $ \langle x(t)\rangle_\infty$ in Fig.~\ref{wp1}. This is
expected as the evaluation of $C_{coh}(\tau)$ involves only the time
behavior of $ \langle x(t)\rangle_\infty$ at two different instants of
time.

Two features of the behavior of $C_{incoh}(\tau)$ are relevant: its
initial value and its decay time.  The initial value of the incoherent
contribution, $C_{incoh}(0)$, is given by the cycle average of the
second cumulant. It has a non-monotonic behavior with $D$. For $D=0.02$,
$C_{incoh}(0)$ is large, consistent with the fact that the second
cumulant at this noise strength is appreciably different from $0$ during
a substantial part of a period.  As the value of $D$ increases, ($D
\le 0.1$), the value of $C_{incoh}(0)$ decreases. This is expected as
the second cumulant is large just during those small time intervals
where most of the forward transitions take place every half-period. For
still larger values of $D$ there are frequent forward and backward jumps
that keep the stationary probability bimodal, and therefore, the initial
value $C_{incoh}(0)$ increases.  For $D=0.02$, the decay of
$C_{incoh}(\tau)$ is very slow, although the decay time is still shorter
than the duration of half a period of the driving force. As $D$
increases, the decay time of the incoherent part becomes shorter. It is
worth to point out that the intra-well noisy dynamics manifests itself
in the behavior of $C_{incoh}(\tau)$. This is most clearly confirmed
noticing the fast initial decay observed in panel d. For smaller values of $D$,
this feature is masked by the long total relaxation time scale, while
for $D=0.2$, the noise strength is so large that there is not a
clear-cut separation between inter and intra-well time scales.

\begin{figure}
\includegraphics[width=10cm]{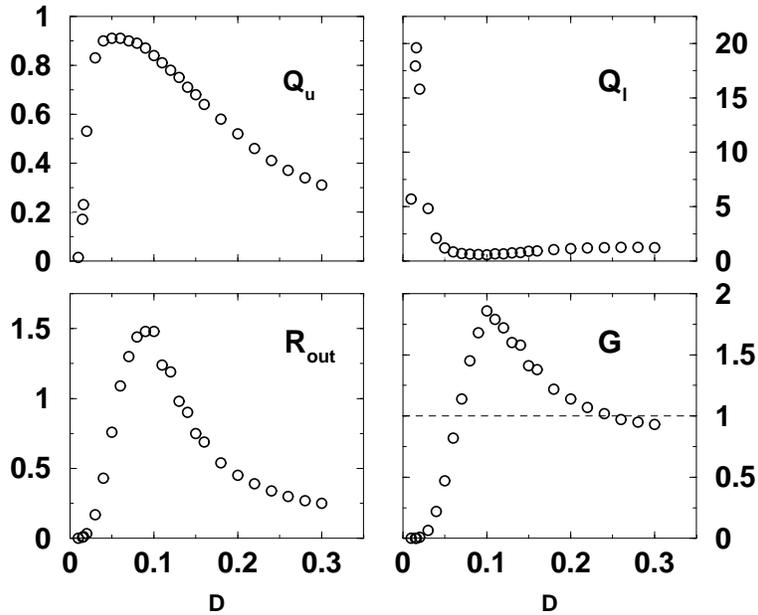}
\caption{\label{wp3} Dependence with $D$ of several SR quantifiers: the
numerator of the SNR ($Q_u$), its denominator ($Q_l$), the output SNR
($R_{out}$) and the SR-gain ($G$) for a rectangular driving force with
duty cycle 1, 
fundamental frequency $\Omega=0.01$ and subthreshold amplitude
$A=0.25$.}
\end{figure}

The above considerations allow us to rationalize the behavior of the
several quantifiers used to characterize SR. Their behaviors with $D$
for $A=0.25$, $\Omega=0.01$ are depicted in Fig.~\ref{wp3}. It should
be noticed that the lowest value of the noise strength used in the
numerical solution of the Langevin equation is $D=0.01$.  For this noise
strength, the values of $Q_u$ and $R_{out}$ are very small, although not
zero. For even lower noise strengths the task becomes computationally
very demanding and expensive, due to the extremely slow decay of the
correlations. For $D$ sufficiently small, however, one does expect $Q_u$
to be larger than $Q_l$, and, consequently, an increase of the numerical
$R_{out}$ as $D$ is lowered.

The quantity $Q_u$ defined in Eq.\ (\ref{num}) depicts a non-monotonic
dependence on $D$ typical of the SR phenomenon.  Its behavior is
expected from the dependence of the amplitude of $C_{coh}(\tau)$ with
$D$ in Fig.~\ref{wp2}.

A non-monotonic behavior with $D$ for the numerically evaluated $Q_l$ is also
observed. The initial value, $C_{incoh}(0)$ and the decay time of
$C_{incoh}(\tau)$ are important in the evaluation of $Q_l$ [see
Eq. (\ref{den})]. For $D=0.01$, the decay time of the incoherent part of
the correlation function is longer than half a period of the driving
force, while for $D=0.02$, it is somewhat shorter than
$T/2$. Consistently with Eq.~(\ref{den}), the value of the integral for
$D=0.01$ is smaller than for $D=0.02$.  As $D$ is further increased, the
influence of the cosine factor in Eq.~(\ref{den}) becomes less important
as the decay time is much shorter than $T/2$.  The drastic fall in the
$Q_l$ values observed for $0.02 < D < 0.1$ is due to the decrease of
$C_{incoh}(0)$ with $D$ (see panels a-c in Fig.~\ref{wp2}) and the
shortening of the decay time. As $D$ is increased further,
$C_{incoh}(0)$ increases and, consequently, $Q_l$ also increases
slightly.

Taking into account the definition of the SNR, [cf. Eq.~(\ref{snr1})],
its behavior with $D$ is not surprising. The numerically obtained SNR
peaks at $D=0.08$, a slightly different value of $D$ from the
one at which $Q_u$ peaks.

\subsection{Anomalous SR-gain behavior for subthreshold driving}
The SR-gain is defined in Eq.\ (\ref{gain}). The numerically determined SR-gain
shows a most interesting feature: we observe a non-monotonic behavior
versus $D$, with values for the gain exceeding unity (!)  for a whole
range of noise strengths. This is strictly forbidden within LRT
\cite{casgom03}; therefore, the fact that the SR-gain can assume values
larger than unity reflects a manifestation of the inadequacy of LRT to
describe the system dynamics for the parameter values considered. 

To rationalize this anomalous SR-gain behavior, we notice that the role of the
noise in the dynamics is twofold. On the one hand, it controls the decay
time of $C_{incoh}(\tau)$. On the other hand, the noise value is
relevant to ellucidate whether the one-time probability distribution is
basically monomodal or bimodal during most of the cycle and,
consequently, it controls the initial value $C_{incoh}(0)$.  As
discussed above, if $D$ is small, the decay time is very large
compared to $t_{asym}$, and the one-time probability distribution is
essentially bimodal. For large values of $D$, the decay of
$C_{incoh}(\tau)$ is fast enough, and the distribution is also bimodal.
The large SR-gain obtained here requires the existence of a range of
intermediate noise values such that: i) $C_{incoh}(\tau)$ decays in a
much shorter time scale than $t_{asym}$ and, ii) the one-time probability
distribution remains monomodal during most of the external cycle.
\subsection{Response to a rectangular input driver with fundamental
frequency $\Omega=0.1$}

As mentioned before, there are several time scales which are important
for the phenomenon of stochastic amplification and gain. In the previous
subsection we have considered an input frequency small enough so that
the inequality $t_{asym} > t_{inter}$ holds for a range of noise values.
Next, we shall analyze the system response to a driving force with
fundamental frequency $\Omega=0.1$, ten times larger than in the
previous case.  We will take the same input amplitude as in the previous
case, $A=0.25$, which is still subthreshold. For this input frequency,
the threshold value for the amplitude strength is determined by
numerically solving the deterministic equation to yield $ A_T \approx
0.39$.

\begin{figure}
\includegraphics[width=10cm]{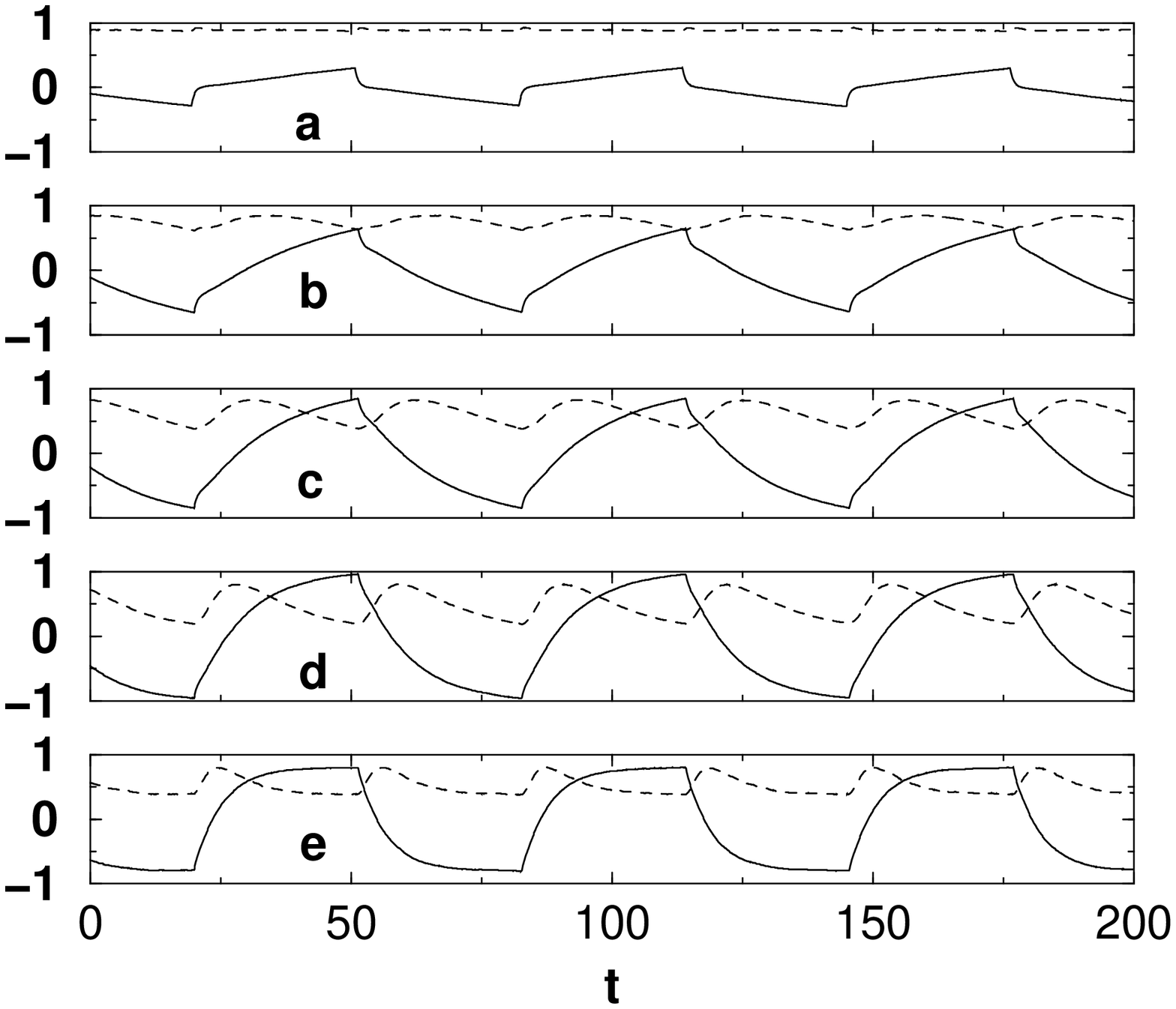}
\caption{\label{wp11} Same as in Fig.~\ref{wp1} but for $\Omega=0.1$}
\end{figure}

The behavior of the first two cumulants for several values of the noise
strength is depicted in Fig.~\ref{wp11} [from top to bottom $D=0.02$
(panel a), $D=0.04$ (panel b), $D=0.06$ (panel c), $D=0.1$ (panel d) and
$D=0.2$ (panel e)]. For all values of $D$, the second cumulant remains
large for most of each half-period.  By contrast with the
lower frequency case, we detect no values of $D$ for which the
probability distribution is monomodal for a significant fraction of each
half period.

\begin{figure}
\includegraphics[width=10cm]{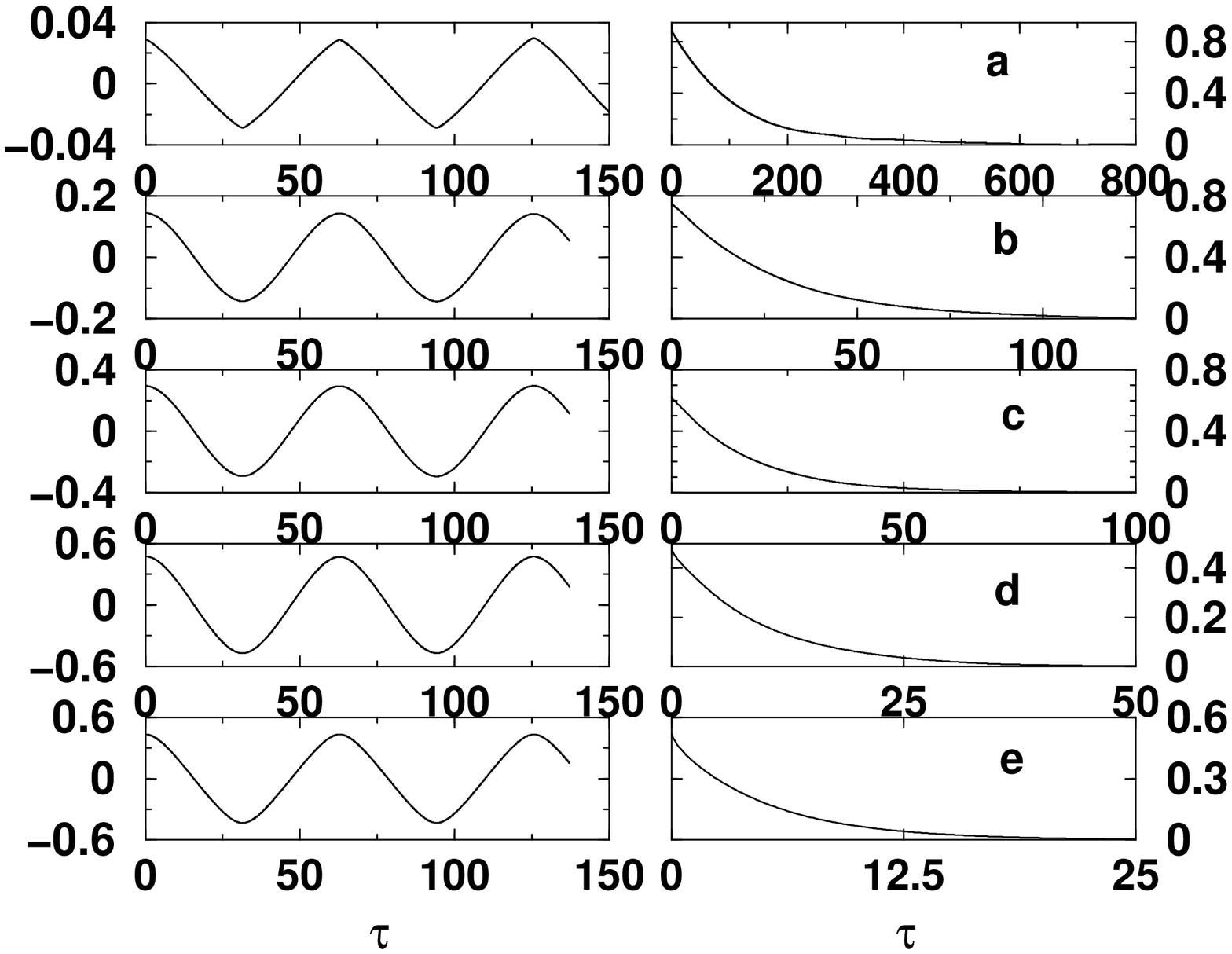}
\caption{\label{wp22} Same as in Fig.~\ref{wp2} but for $\Omega=0.1$.}
\end{figure}

In Fig.~\ref{wp22} the behavior of the coherent (left panels) and
incoherent (right panels) parts of the correlation function are
presented for (from top to bottom) $D=0.02$ (panel a), $D=0.04$ (panel
b), $D=0.06$ (panel c), $D=0.1$ (panel d) and $D=0.2$ (panel e). The
amplitude of the coherent oscillations shows a nonmonotonic behavior
with $D$. The incoherent part has initial values which remain very large
in comparison with the corresponding ones for $\Omega=0.01$ (compare
with Fig.~\ref{wp2}), consistently with the large value of the second
cumulant.  The decay times are roughly the same for both frequencies.

\begin{figure}
\includegraphics[width=10cm]{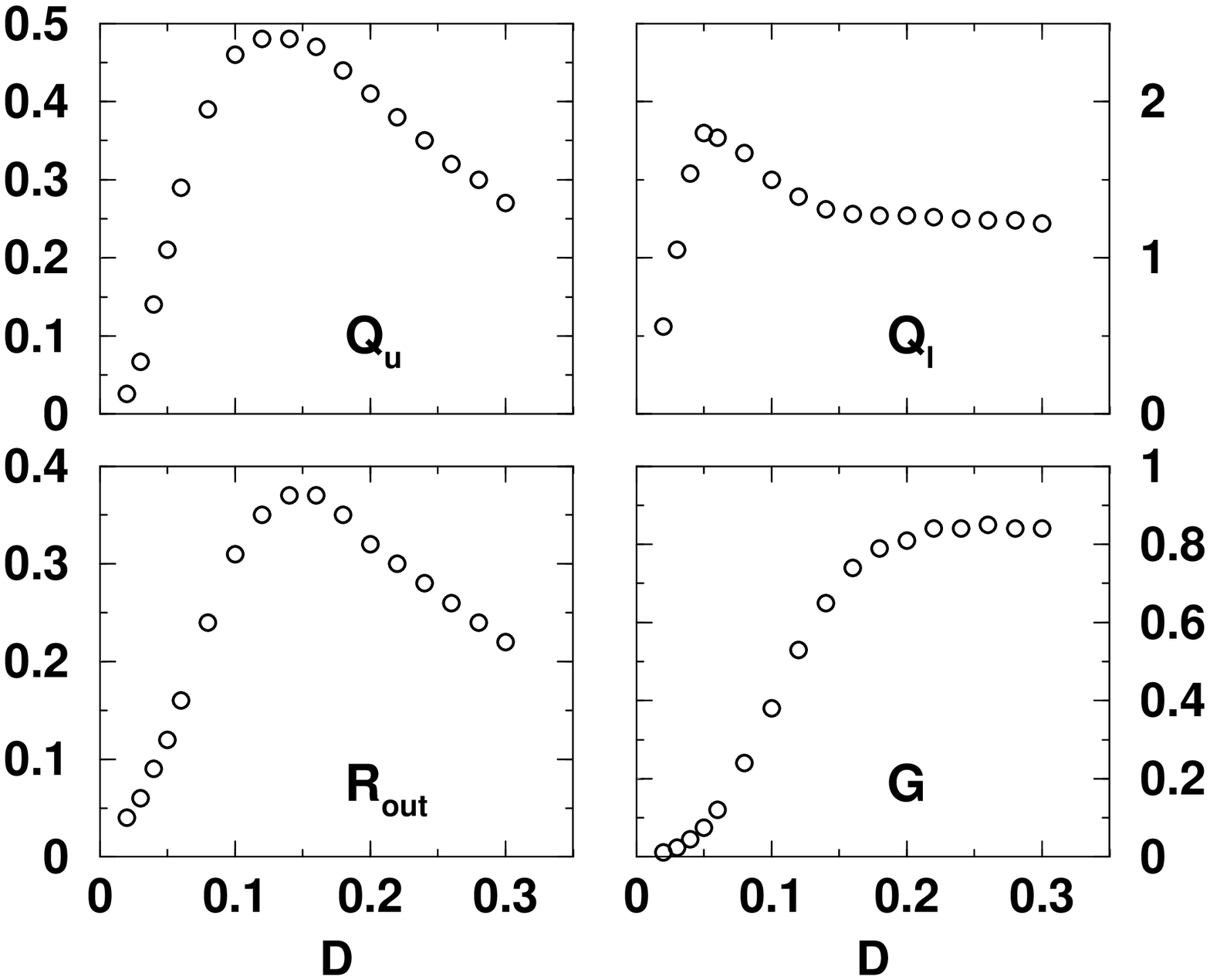}
\caption{\label{wp33} Same as in Fig.~\ref{wp3} but for $\Omega=0.1$.}
\end{figure}

In Fig.~\ref{wp33} we show the behavior of the several SR quantifiers as
a function of $D$.  The comparison of Figs.~\ref{wp3} and \ref{wp33}
indicates that $Q_u$ , $Q_l$ and $R_{out}$ have the same qualitative
behavior for both frequencies. The non-monotonic dependence on $D$ of
$Q_u$ and $R_{out}$ are indicative of the existence of SR (for both
frequencies) for the subthreshold input amplitude and in the ranges of
$D$ values considered. The most relevant quantitative difference is that
for $\Omega=0.1$ the SR-gain remains less than unity.

\section{Conclusions}
With this work, we have analyzed the phenomenon of SR within the context
of a noisy, bistable symmetric system driven by time periodic,
rectangular forcing possessing a duty cycle of unity. The numerical
solution of the Langevin equation allows us to analyze the long-time
behavior of the average, the second cumulant and the coherent and
incoherent parts of the correlation function. For subthreshold input
signals we determined the SNR, together with its numerator and
denominator evaluated separately, for a wide range of noise strengths
$D$.

As a main result we find the simultaneous existence of a typical
non-monotonic behavior versus the noise strength $D$ of several
quantifiers associated to SR; in particular SR-gains larger than unity
are possible for a subthreshold rectangular forcing possessing a duty
cycle of unity.  This finding is at variance with the recent claim in
Refs.~\cite{ginmak01,makging02} that pulse-like signals with small duty
cycles are needed to obtain SR-gains larger than unity. This unexpected
result occurs indeed whenever the inequality $t_{asym} > t_{inter}$
holds.  This is most easily achieved with low frequency inputs. As the
input frequency increases, that inequality is not satisfied for
sufficiently small values of $D$: even though SR then still exists, it
is not accompanied by SR-gains exceeding unity.  Furthermore, in
Refs.~\cite{ginmak01,makging02}, SR-gains larger than unity are only
obtained with input amplitudes larger than $0.8\,A_T$. By contrast, in
this work, we have shown that such a large value for the input amplitude
is not needed (we have used $A\approx 0.68\,A_T$).

The simultaneous occurrance of SR and SR-gains larger than unity is
associated to the fact that, for some range of noise values, the decay
time of the incoherent part of the correlation function is much shorter
than $t_{asym}$ and also the probability distribution is basically monomodal
during most of the cycle of the driving force.

\begin{acknowledgments}
We acknowledge the support of the Direcci\'on General de
Ense\~nanza Superior of Spain (BFM2002-03822), the Junta de
Andaluc\'{\i}a, the DAAD program "Acciones Integradas" (P.H., M.
M.) and the Sonderforschungsbereich 486 (project A10) of the Deutsche
Forschungsgemeinschaft.
\end{acknowledgments}

\newpage

\begin{thebibliography}{100}

\bibitem{PT96}
A. R. Bulsara and L. Gammaitoni, Physics Today {\bf 49}, No. 3, 39
(1996).
\bibitem{RMP}
L. Gammaitoni, P. H\"anggi, P. Jung, and F. Marchesoni, Rev. Mod.
Phys. {\bf 70}, 223 (1998).
\bibitem{Wiesenfeld98}
K. Wiesenfeld and F. Jaramillo, Chaos {\bf 8}, 539 (1998).
\bibitem{Anish99}
V. S. Anishchenko, A. B. Neiman, F. Moss, and L. Schimansky-Geier,
Usp. Fiz. Nauk {\bf 169}, 7  (1999).
\bibitem{Chemphyschem02}
P. H\"anggi, CHEMPHYSCHEM {\bf 3}, 285 (2002).
\bibitem{McNWie89} B. McNamara and K. Wiesenfeld,  Phys. Rev. A {\bf
39}, 4854 (1989).
\bibitem{Hanggi78}
P. H\"anggi, Helv. Phys. Acta, {\bf 51}, 202 (1978).
\bibitem{PR82} P. H\"anggi, and H. Thomas, Phys. Rep. {\bf 88}, 207 (1982).
\bibitem{EPL89} P. Jung and P. H\"anggi, Europhys. Lett. {\bf 8}, 505 (1989).
\bibitem {GM89} L. Gammaitoni, E. Menichella-Saetta, S. Santucci,
F. Marchesoni, and C. Presilla, Phys. Rev. A {\bf 40}, 2114 (1989).
\bibitem{PRA91}
P. Jung and P. H\"anggi, Phys. Rev.  A {\bf 44}, 8032 (1991).
\bibitem{EPLcasado}
J. Casado-Pascual, J. G\'omez-Ord\'o\~nez, M. Morillo, and P. H\"anggi,
Europhys. Lett. {\bf 58}, 342 (2002).
\bibitem{FNLcasado}
J. Casado-Pascual, J. G\'omez-Ord\'o\~nez, M. Morillo, and P. H\"anggi,
Fluct. Noise Lett. {\bf 2}, L127 (2002).
\bibitem{dykman92} M. I. Dykman,
R. Mannella, P. V. E. McClintock, and N. G. Stocks,
Phys. Rev. Lett. {\bf 68}, 2985 (1992).
\bibitem{junhan93} P. Jung and P. H\"anggi,  Z.  Physik B {\bf
90}, 255 (1993).
\bibitem{dewbia95} M. DeWeese and W. Bialek, Il Nuovo Cimento
{\bf17D}, 733 (1995).
\bibitem{casgom03} Jes\'us Casado-Pascual, Claus Denk, Jos\'e
G\'omez-Ord\'o\~nez, Manuel Morillo, and Peter H\"anggi, Phys. Rev. E
{\bf 67}, 036109 (2003).
\bibitem{haninc00} P. H\"anggi, M. Inchiosa, D. Fogliatti and
A. R. Bulsara, Phys. Rev. E {\bf 62}, 6155 (2000).
\bibitem{ginmak00} Z. Gingl, R. Vajtai, and P. Makra, in ``Noise
in Physical Systems and 1/f Fluctuations '', ICNF 2001,  G.
Bosman, editor (World Scientific, 2002), pp. 545-548.
\bibitem{ginmak01} Z. Gingl, P. Makra, and R. Vajtai,  Fluct.
Noise Lett. {\bf 1}, L181, (2001).
\bibitem{Morillo95}
M. Morillo and J. G\'omez-Ord\'o\~nez, Phys. Rev. E {\bf 51}, 999
(1995).
\bibitem{hel79} E. Helfand, Bell Sci. Tech. J., {\bf 58}, 2289 (1979).
\bibitem{grehel81} H. S. Greenside and E. Helfand, Bell Sci. Tech. J.,
{\bf 60}, 1927 (1981).
\bibitem{makging02} P. Makra, Z. Gingl and L. B. Kish, Fluct.
Noise Lett. {\bf 1}, L147, (2002).
\end{thebibliography}
\end{document}